\documentclass[a4paper,11pt]{article}
\pdfoutput=1 

\usepackage{jheppub} 
\makeatletter
\DeclareRobustCommand*{\bfseries}{%
  \not@math@alphabet\bfseries\mathbf
  \fontseries\bfdefault\selectfont
  \boldmath
}
\makeatother

\usepackage{calc}
\usepackage{rotating}
\usepackage[english]{babel}
\usepackage{graphicx}
\usepackage{subfig}
\usepackage{float}
\usepackage{amsmath}
\usepackage{amssymb}
\usepackage{amsthm}
\usepackage{latexsym}
\usepackage{dcolumn}
\usepackage{hyperref}
\input macros.tex

\nc\fb{\ensuremath{{\bar 5}} {}}
\nc\fby{\ensuremath{{\overline 5}_Y} {}}
\nc\fy{\ensuremath{{5}_Y} {}}
\nc\upq{\ensuremath{U(1)_{\textrm{PQ}}}}
\nc{\qpq}{q}
\nc\lsusy{\ensuremath{\La} {}}
\nc\msusy[1][]{\ensuremath{M_{\textrm{SUSY}}}}
\nc\my{\ensuremath{M_Y} {}}
\definecolor{gray}{rgb}{.5,.5,.5}

\restylefloat{figure}

\title{Flavored gauge mediation in the Peccei-Quinn NMSSM}

\author{Kamila Kowalska,$^{a}$}
\author{Jacek Pawe{\l}czyk$^{b}$}
\author{and Enrico Maria Sessolo$^{a}$}

\affiliation{$^{a}$National Centre for Nuclear Research,\\
\textrm{\,}Ho{\. z}a 69, 00-681 Warsaw, Poland\\
$^{b}$Institute of Theoretical Physics, University of Warsaw,\\ 
\textrm{\,}Pasteura 5, 02-093 Warsaw, Poland} 

\emailAdd{kamila.kowalska@ncbj.gov.pl}
\emailAdd{jacek.pawelczyk@fuw.edu.pl}
\emailAdd{enrico.sessolo@ncbj.gov.pl}

\abstract{We investigate a particular version of the Peccei-Quinn (PQ) NMSSM characterized 
by an economical and rigidly hierarchical flavor structure and based on flavored gauge mediation 
and on some considerations inspired by string theory GUTs.
In this way we can express the Lagrangian of the PQ NMSSM through very few parameters. 
The obtained model is studied 
numerically and confronted with the most relevant phenomenological constraints. 
We show that typical spectra are for the most part too heavy to be significantly probed at the LHC, 
but regions of the parameter space exist yielding signatures that might possibly be observed during Run II.
We also calculate the fine tuning of the model. We show that, in spite of the appearance of large scales 
in the superpotential and soft terms, it does not exceed the tuning present in the MSSM for equivalent spectra,
which is of the order of $10^4$.}

\begin{document}
\maketitle
\section{\label{sec:intro}Introduction}

Low scale supersymmetry (SUSY) is still a good candidate for physics beyond the Standard Model (SM) despite a discouraging 
lack of positive signals in the first run of the LHC. 
In its uncostrained version the Minimal Supersymmetric Standard Model (MSSM) does not provide a solution to the origin of the flavor structure
of the Yukawa couplings, but hierarchical textures are typical of many UV completions,  
starting from the seminal paper of Froggatt and Nielsen~\cite{Froggatt:1978nt}, 
up to more recent developments in  
F-theory Grand Unified Theories (GUTs), see~\cite{Vafa:1996xn,Donagi:2008ca,Beasley:2008dc,Beasley:2008kw,Heckman:2009mn,Cecotti:2009zf,Heckman:2010bq}
for some early papers. 
Interestingly, F-theory GUTs provide 
a natural connection to SUSY breaking, as the visible chiral matter as well as the messengers 
responsible for conveying the breaking to the visible sector 
originate from the same D7-brane intersection~\cite{Heckman:2009mn}. A consequence of this fact is the requirement that 
matter and messengers share common Yukawa couplings. 

From the phenomenological point of view, a connection between flavor and SUSY breaking has been developed in models with 
flavored gauge mediation~\cite{Dine:1996xk,Giudice:1997ni,Chacko:2001km,Chacko:2002et},
a branch of gauge mediated SUSY breaking (GMSB)~\cite{Dine:1981gu,Nappi:1982hm,AlvarezGaume:1981wy,Dine:1993yw,Dine:1994vc,Dine:1995ag,Giudice:1998bp} 
that links the flavor and messenger sectors of the MSSM. 
Along these lines, one of us recently proposed \cite{Pawelczyk:2013tza} a very economical model that successfully 
incorporates the advantages of Froggat-Nielsen-like hierarchical Yukawa structures into the GMSB messenger sector.   

In \cite{Pawelczyk:2013tza}, all the soft terms depend on one extra parameter, $h_3$, 
which is the coupling between the messenger in the \textbf{5} representation of $SU(5)$ and chiral matter. 
Through well known mixing effects, the extra coupling can also produce a large third-generation soft trilinear term, $A_t$, 
which can be used to enhance the radiative corrections to the Higgs mass.
A phenomenological analysis of the low-energy limit of this minimal model in the MSSM 
was presented in~\cite{Jelinski:2014uba}. It was shown that, despite quite restrictive constraints on the nature of the matter-messenger couplings, 
which arise from the flavor structure, it is easy to find regions of the parameter space characterized by particle spectra 
consistent with the bounds from direct searches for squarks and gluinos at the LHC, and 
with the measured value of the Higgs boson mass, $m_h\simeq 125\gev$, as well as a number of constraints from 
flavor-changing neutral current (FCNC) in low-energy processes. 

However, the matter/messenger system of Ref.~\cite{Pawelczyk:2013tza} is fully consistent with hierarchical flavor structures when the ratio
of the Higgs doublets' vacuum expectation values (vev's), \tanb, is of the order of a few and not larger. 
As is well known, in the MSSM small values of \tanb\ reduce the size of the tree-level Higgs mass, which therefore requires 
substantial radiative corrections that imply large soft masses and a consequently high
level of fine tuning.
Thus, in this paper we analyze whether it is possible to obtain a low-energy limit of the model presented in \cite{Pawelczyk:2013tza}
in the Next-to-Minimal Supersymmetric Standard Model (NMSSM).
As is well known, this is achieved by introducing one additional gauge singlet chiral field, $S$, that couples to the Higgs
sector in the superpotential. Because the tree-level value of the Higgs mass for low \tanb\ values can be easily made larger than 
in the MSSM, the NMSSM seems to be a more natural choice as a low-energy effective theory originating from the model in~\cite{Pawelczyk:2013tza}.  

In particular, we investigate here the Peccei-Quinn (PQ) version of the NMSSM, which is characterized by the vanishing of the $S^3$ coupling, 
$\k=0$~\cite{Fayet:1974fj,Fayet:1974pd,Fayet:1976et,Hall:2004qd,Feldstein:2004xi,Barbieri:2007tu,Gherghetta:2012gb}. 
We will build the PQ NMSSM effective action based on arguments from its possible UV completion.
It is interesting that similar models appeared in F-theory constructions~\cite{Heckman:2009mn}, so that we will often invoke 
F-theoretic arguments to justify our assumptions.

The purpose of the paper is twofold.
We construct a version of the PQ NMSSM, based on a string inspired UV completion and on flavored GMSB, 
which yields a very predictive framework. 
The effective action is defined by just a handful of free parameters and we show that the obtained spectra are consistent
with the basic phenomenological constraints, although we also show that they are for the most part too heavy to be significantly explored at the LHC.

We further show that the PQ NMSSM analyzed here requires the unavoidable introduction of nonstandard tadpole superpotential and Lagrangian terms, 
which are linear in the $S$ field, and terms quadratic in $S$ as well, all of them characterized by typical scales that are by orders of magnitude higher
than the SUSY breaking scale.
Unlike the standard GMSB soft terms, these new terms break the global symmetry of 
the PQ NMSSM, \upq, whose existence is at the origin of the $\k=0$ choice. In the string theory setup, 
\upq\ is a remnant of a local gauge symmetry which has been spontaneously broken.
As is well known, the corresponding gauge boson acquires a large (GUT scale) mass through the Green-Schwarz mechanism~\cite{Green:1984sg}.
Here we break \upq\ through the vev of a scalar field $X$, which will also play the additional role of the SUSY breaking spurion.

On the other hand, we show that despite the presence of these extra large-scale terms the model does \textit{not} 
present fine-tuning levels higher than those already present in its MSSM version, i.e., $\sim 10^4$. 
This is due to a cancellation among some specific terms entering the fine-tuning measure, 
because of relations determined by our UV completion.
Thus, this model is partially immune from the mild 
``tadpole problem''~\cite{Ferrara:1982ke,Polchinski:1982an,Nilles:1982mp,Lahanas:1982bk,Ellwanger:1983mg,Bagger:1993ji,Jain:1994tk,Bagger:1995ay} 
of the General NMSSM~\cite{Ellwanger:2008py}, according to which the extra terms dramatically amplify the  
effective theory's sensitivity to the high energy physics.  

The paper is organized as follows. In \refsec{sec:1} we introduce the model. We start from the 
flavored GMSB structure of the superpotential and progress subsequently to introducing one by one the additional 
terms that will define our version of the PQ NMSSM. In \refsec{sec:pheno} we perform a numerical analysis of the model. We
identify the regions of the parameter space consistent with the constraints from the Higgs mass measurement and LHC searches,
and we provide some benchmark points useful for the discussion and also possible collider signatures.
In \refsec{sec:fine-tuning} we discuss the fine tuning of the model, and prove that it is not larger than 
the present level found in the MSSM, despite the presence of tadpole and quadratic terms in the singlet field.
We finally provide our summary and concluding remarks in \refsec{sec:summary}.

Additionally, we provide four appendices to the text, dedicated, respectively, to
the GMSB calculation of the soft masses in the model; to the explicit calculation of the \upq\ 
breaking effective terms of the superpotential and soft Lagrangian; to the explicit estimate of the size of our constants;
and to the explicit form of the tree-level Higgs mass in the PQ NMSSM.

\section{The model}\label{sec:1}

\subsection{Flavored GMSB with hierarchical Yukawa couplings}

A simple model combining the advantages of F-theory model building and 
flavored GMSB was introduced in~\cite{Pawelczyk:2013tza}. 
Since the visible matter and the messengers have the same origin they should also present a common hierarchical structure of their couplings, e.g., 
a structure of the Froggatt-Nielsen type \cite{Froggatt:1978nt}. 
This reasoning basically implies that the messengers are in \mf\ representations of $SU(5)$. 
The simplest construction of this type contains three chiral families of $\mt$s, 
four $\mfb$s and one extra multiplet $\mf_Y$, which is necessary to form a vector pair of messengers and cancel the resulting anomalies. 

The relevant superpotential terms using $SU(5)$ representations at the GUT scale are
\beqa\label{mainY}
W\supset\sum_{i,j}y^u_{ij}\mt_i \mt_j (\mf_H)_2+
\sum_{i,J}\widehat{y}_{iJ}\mt_i \mfb_J (\mfb_H)_2
+(\sum_J a_J \mfb_J) \mf_Y X\,,
\eeqa
where in Eq.~(\ref{mainY}) flavor indices run as $i,j=1,2,3$ and $J=1,2,3,4$, and the subscript ``2'' attached to the Higgs 
fields highlights that we take into account only the doublet part of the $\mathbf{5}_H$ and $\mathbf{\bar{5}}_H$ Higgs multiplets.
In this regard, we recall here that in F-theory GUTs doublet-triplet splitting can be obtained through an appropriate choice of internal fluxes (see, e.g., 
Secs.~10.3 and 12 of Ref.~\cite{Beasley:2008kw}).
 
We assume that all the couplings in Eq.~(\ref{mainY}) have a hierarchical structure, i.e.,
$y^u_{i\,j}\ll y^u_{i+1\,j}$\,, $y^u_{i\,j}\ll y^u_{i\,j+1}$ for all $i,j$ (and similarly for the $\widehat{y}_{iJ}$), 
and also $a_J\ll a_{J+1}$.
One can claim~\cite{Cecotti:2009zf} that the couplings with top-most indices, $y^u_{33},\ \widehat {y}_{34}$, and $a_4$ are 
all of the same order. The measured value of the top quark mass sets  $y_t=y^u_{33}$ to be of 
$\mathcal{O}(1)$ if the renormalization group (RG) flow does not change $y_t$ drastically.
Note also that, as was mentioned in \refsec{sec:intro}, the hierarchy of the couplings is 
better enforced for values of \tanb\ not much larger than 1, as $\widehat{y}_{33}/\widehat{y}_{34}\sim\tanb\,m_b/m_t$\,.

The spurion $X$ acquires a vacuum expectation value (vev), $\vev X$, which gives a mass, $\my=\vev X$, 
to the messengers $\mf_Y$ and $\mfb_Y$, where $\mfb_Y=\sum_J a_J \mfb_J$. The other $\mfb$s remain massless.
The spurion also triggers SUSY breaking through its nontrivial $F$-term: 
$\widehat{X}\rightarrow\vev X+\theta^2 F_X$\,.
We assume here that the dynamics of the spurion is governed by the physics of the high scale, so that 
it is effectively a non-dynamical field below the messenger scale.

After diagonalizing the mass matrices in Eq.~(\ref{mainY}),  
we can write down the terms with $\mathcal{O}(1)$ couplings that will be relevant for our analysis:
\beqa\label{leadY}
W\supset y_t \mt_3\mt_3 (\mf_H)_2+ h_3 \mt_3\mfb_Y(\mfb_H)_2+\mfb_Y \mf_Y X\,,
\eeqa
where $y_t\approx 1,\ h_3\equiv \widehat{y}_{34}\approx 1$.
By integrating out the messengers one can obtain the soft terms of the MSSM explicitly displayed in Appendix~\ref{softs}.

\subsection{NMSSM GMSB with PQ symmetry\label{sec:gmsbpq}}

In the NMSSM one introduces an extra chiral superfield $S$, whose vev generates the $\mu$ term at the electroweak (EW) scale. 
We consider here the hypothesis that the fields of the model are charged under a PQ symmetry, 
\upq,
which introduces additional restrictions on the possible couplings. In particular, as we shall see below, it forbids any superpotential term for $S$ alone ($\k=0$).  
We also assume that the PQ charges of the $X$ and $S$ fields are opposite in sign, $\qpq(S)=-\qpq(X)$ (as considered, e.g., in~\cite{Heckman:2009mn}), 
which leads to certain necessary terms in the Lagrangian. 
Incidentally, \upq\ also helps prevent 
fast baryon decay due to GUT processes \cite{Pawelczyk:2010xh}.

The resulting \upq\ charges are summarized in the following table:
\beqa\label{model1}
\begin{tabular}
[c]{|c|c|c|c|c|c|c|c|}
\hline
& $\mt_{1,2,3}$ & $\mfb_{1,2,3},\, \mfb_Y$& $\mf_{H}$ & $\mfb_{H}$& $\mf_Y$ &
$S$& $X$ \\\hline
$\qpq$ & $+1/4$ & $+1/4$  & $-1/2$ & $-1/2$ &$+3/4$ & $+1$ &$-1$ \\\hline
\end{tabular}
\eeqa
where the charges are normalized in such a way that the largest one equals $1$.
Note that this implies that the spurion $X$ does not couple to the Higgs fields, 
while $S$ cannot couple to the messengers in dimension 4 operators. Moreover,
as mentioned above, there cannot exist polynomial couplings in $S$ and $X$ alone so that, e.g., $\k=0$ 
in the notation of the usual $\mathbb{Z}^3$-symmetric NMSSM \cite{Ellwanger:2009dp}. 

In summary, the only renormalizable superpotential term preserving \upq\ and containing $S$ 
is\footnote{We drop here the mass term $XS$, which is unnatural in F-theoretic constructions, see Appendix~\ref{app:eff}.} 
\beq\label{lam}
W_S=\la S H_u H_d\,,
\eeq
which after SUSY breaking gives rise to a trilinear soft Lagrangian contribution, $A_\lam$, and a scalar soft mass for the singlet, $m_S^2$, 
\beq
-L_{S,\textrm{soft}}=\lam A_\lam H_u H_d S+m_S^2|S|^2\,,\label{mes}
\eeq
whose explicit expression in terms of the GMSB parameters is given in Appendix~\ref{softs}.  

As one can easily see, \upq\ is anomalous. 
It is known that the anomaly can be canceled by the Green-Schwartz mechanism \cite{Green:1984sg}, which gives a mass to the \upq\ gauge boson of the order of \lgut\,.

Note that \vev X breaks \upq, so that at low energies there appear additional 
effective \upq\ breaking interaction terms. 
We present them in the next subsection and discuss them in detail in Appendix~\ref{app:eff}.

\subsection{Effective interactions below the messenger scale\label{sec:eff}}

In addition to \refeq{lam} and \refeq{mes}, other contributions to the superpotential and soft Lagrangian might arise below $M_Y$. 
As we explain in Appendix~\ref{app:eff}, these can be due, for example, to instanton effects, exchange of heavy chiral \upq\ neutral superfields, 
exchange of heavy gauge bosons, or Giudice-Masiero~\cite{Giudice:1988yz} effective terms in the K\"{a}hler potential.
The most important terms for the low-energy dynamics will be those containing powers of the spurion superfield,  
$X$, because $X$ receives a large vev. 

Additional terms to the superpotential and soft SUSY-breaking Lagrangian take the form
\beqa
\d W_S &=&\xi_F S+\frac{1}{2}\mu'S^2\,,\label{superpot}\\
-\d L_{S,\textrm{soft}}&=& \delta m_S^2|S|^2
+\left(\frac{1}{2}m_S'^2 S^2+\xi_S S+\textrm{h. c.}\right)\label{softlag}
\,,
\eeqa
where we have used the standard notation of \cite{Ellwanger:2009dp} and we do not explicitly distinguish
between the superfields and their scalar components. 
Importantly, to a very good approximation the relations 
\beq
\kappa\approx0\,\textrm{ and}\,\,\,A_\kappa\approx 0\,
\eeq
still hold, even while there are other massive couplings that effectively break \upq. 

The superpotential and Lagrangian elements in Eqs.~(\ref{superpot}) and (\ref{softlag}) are best expressed in terms
of $x=M_Y/\lgut$ and $\Lambda=F_X/M_Y$. 
They are given by (see Appendix~\ref{app:eff} for full details)
\beqa
\delta m_S^2&=&-|\la_\Psi|^2|\lsusy|^2 x^2\,, \non
\xi_F&=&I_{1,3} \lgut^2\,x^2,\non
\mu'&=&-\la_\Psi^2 \lgut\ x^2,\non
\xi_S&=&-2I_{1,3}\lsusy \lgut^2\,x^2 =-2 \xi_F\Lambda\,, \non
m_S'^{2}&=& 2\la_\Psi^2\lsusy\lgut\ x^2=-2\mu' \lsusy\,,\label{eff_model}
\eeqa
where $I_{1,3}$ is a dimensionless instanton contribution, 
and $\lambda_{\Psi}$ is a $\mathcal{O}(1)$ ``Yukawa'' coupling between the
$X$ and $S$ superfields and a heavy chiral \upq\ neutral superfield, $\Psi$, that is integrated out below \lgut\,.
Without loss of generality, in what follows we fix $\lgut= 2\times 10^{16}\gev$ and $\lam_{\Psi}= 1$\,. 

All the soft terms and superpotential parameters in Eqs.~(\ref{eff_model}) are defined at \my and then 
renormalized to lower energy through RG flow. 
The model is therefore described entirely by 
5 free parameters defined at the messenger scale: 
\begin{equation}\label{pars}
M_Y\,(\textrm{or }x), \Lambda, h_3, \lambda, \xi_F\,,
\end{equation} 
where we have used the more familiar superpotential tadpole $\xi_F$ in place of the numerical instanton coefficient $I_{1,3}$\,. 
As will be clear below, from a phenomenological point of view one is interested in a somewhat limited range of $\La$, 
so that the number of really free parameters can then be reduced to 4, of which only 2, $x$ and $\xi_F$,  
set the overall scale of the additional terms in Eqs.~(\ref{superpot})-(\ref{softlag}).

We want to highlight here that
\beq\label{ratio}
\frac{\xi_S}{\xi_F}=\frac{m_S'^2}{\mu'}\,,
\eeq
which holds up to leading terms (see Appendix~\ref{estimates}). Equation~\refeq{ratio}
is RG invariant after suppressing subdominant terms in the RG equations. 
This characteristic will turn out to be essential to reducing the fine tuning of the model by approximately two orders of magnitude, 
thus largely ameliorating the tadpole problem, 
as we discuss in detail in Sec.~\ref{sec:fine-tuning}.

\section{Phenomenological analysis \label{sec:pheno}}

In this sections we present the results of our numerical analysis, which determines the phenomenological constraints 
on the parameter space of the model and possible signatures at present and future experiments.
In practice, the most important constraints come from the measurement of the Higgs boson
mass, $m_{h_1}\approx 125\gev$~\cite{Chatrchyan:2012ufa,Aad:2012tfa} and the bounds from direct searches for SUSY at the LHC, 
of which the most relevant are the ones on the gluino mass, $m_{\widetilde{g}}\gtrsim 1.5\times 10^3\gev$~\cite{Aad:2014wea,Aad:2015hea,Khachatryan:2015vra},
and on heavy stable charged particles~\cite{Chatrchyan:2013oca}.  

We calculate spectra and NMSSM parameters with \texttt{NMSSMTools v4.5.1}~\cite{NMSSMTools}, 
which we modified to incorporate the $h_3$-dependent and $\lam$-dependent contributions 
to the soft terms given in Eqs.~\refeq{gmsbsoft} of Appendix~\ref{softs}, and the additional masses and tadpoles defined in Eqs.~(\ref{eff_model}).

To guide the scanning procedure we use \texttt{MultiNest}~\cite{Feroz:2008xx}. 
We scan with flat priors in the parameters $\Lambda$, $M_Y$, $\lam(\msusy)$, $h_3$, and $\tanb$. 
The fundamental parameter $\lam(M_Y)$ has been traded 
for the renormalized value of $\lam$ at the scale of the geometrical average of the 
stop masses, \msusy; the fundamental parameter $\xi_F$ has been traded for the ratio of the Higgs doublets' vevs, \tanb, through the electroweak 
symmetry breaking (EWSB) conditions. 

We scan in the following ranges:
\begin{equation}
10^5\gev\leq\Lambda\leq 10^6\gev\,,\,\,\,\,\,\,\,\,10^9\gev\leq M_Y\leq 10^{14}\gev\,,\nonumber
\end{equation}
\begin{equation}\label{h3range}
0.1\leq h_3\leq 1\,,\,\,\,\,\,\,\,\,0.1\leq \lam(\msusy)\leq 2\,,\,\,\,\,\,\,\,\,\,1\leq\tanb\leq 8\,.
\end{equation}

As was mentioned above, the LHC lower bound on the gluino mass affects the choice of scanning range for $\Lambda$, 
which we constrain to a narrow interval, as the soft masses of the gluino and the bino strongly depend on this parameter.
Below $\Lambda\approx 10^5\gev$ the gluino tends to become too light, while for $\Lambda>10^6\gev$ 
the model becomes substantially and uncomfortably more fine-tuned.
The lower bound on $M_Y$ comes from requiring the two-loop GMSB expression of Eqs.~\refeq{gmsbsoft} to be the dominant contribution to the soft masses,
while the upper bound comes from the requirement of being well within the range of validity of the 
effective theory below \lgut\,: for $M_Y>10^{14}\gev$, the first-order expansion at the 
origin of Eqs.~\refeq{eff_model} is not well-justified.
Note finally, that we allow for a wide range of $\lambda$ values, to counterbalance our choice of low \tanb. 

In addition to the fundamental parameters we scan in some of the SM nuisance parameters: the strong coupling constant, 
the $\overline{MS}$ value of the bottom quark mass, 
and the top quark pole mass, which we include in the likelihood function. For these we adopt normal distributions based on the most recent PDG~\cite{Agashe:2014kda} 
central values and experimental uncertainties.
The scans are driven by a Gaussian likelihood function for the experimental measurement of the Higgs mass,
where we added to the experimental uncertainties, in quadrature, a theoretical uncertainty of approximately 3\gev.
This choice is motivated by the large uncertainty still present in the calculation of the Higgs mass, which includes the choice 
of renormalization scheme, missing higher order contributions, and numerical differences between various existing public codes
(see~\cite{Staub:2015aea} for a recent discussion).  

\begin{figure}[t]
\centering
\subfloat[]{%
\label{fig:a}%
\includegraphics[width=0.50\textwidth]{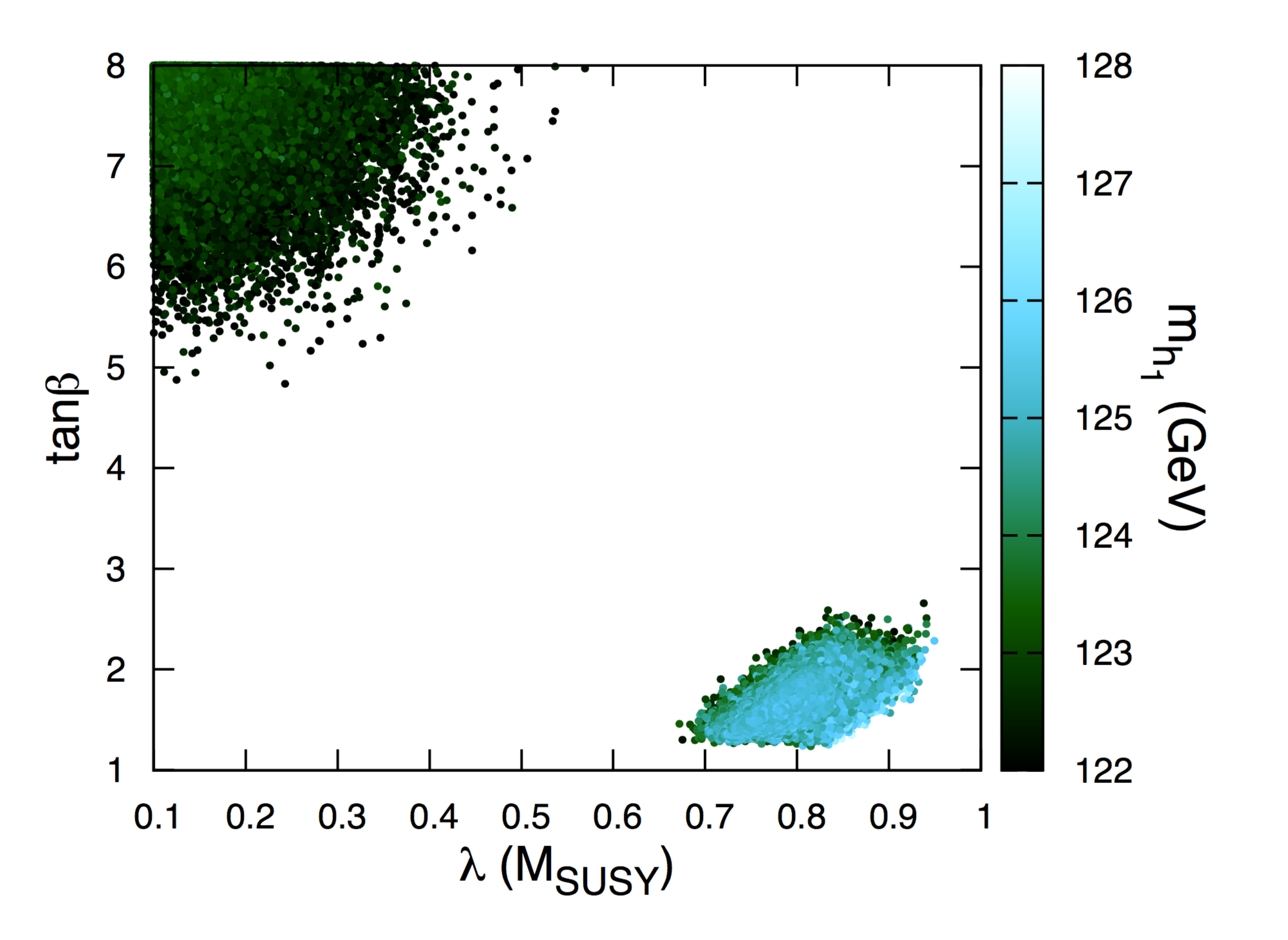}
}%
\subfloat[]{%
\label{fig:b}%
\includegraphics[width=0.50\textwidth]{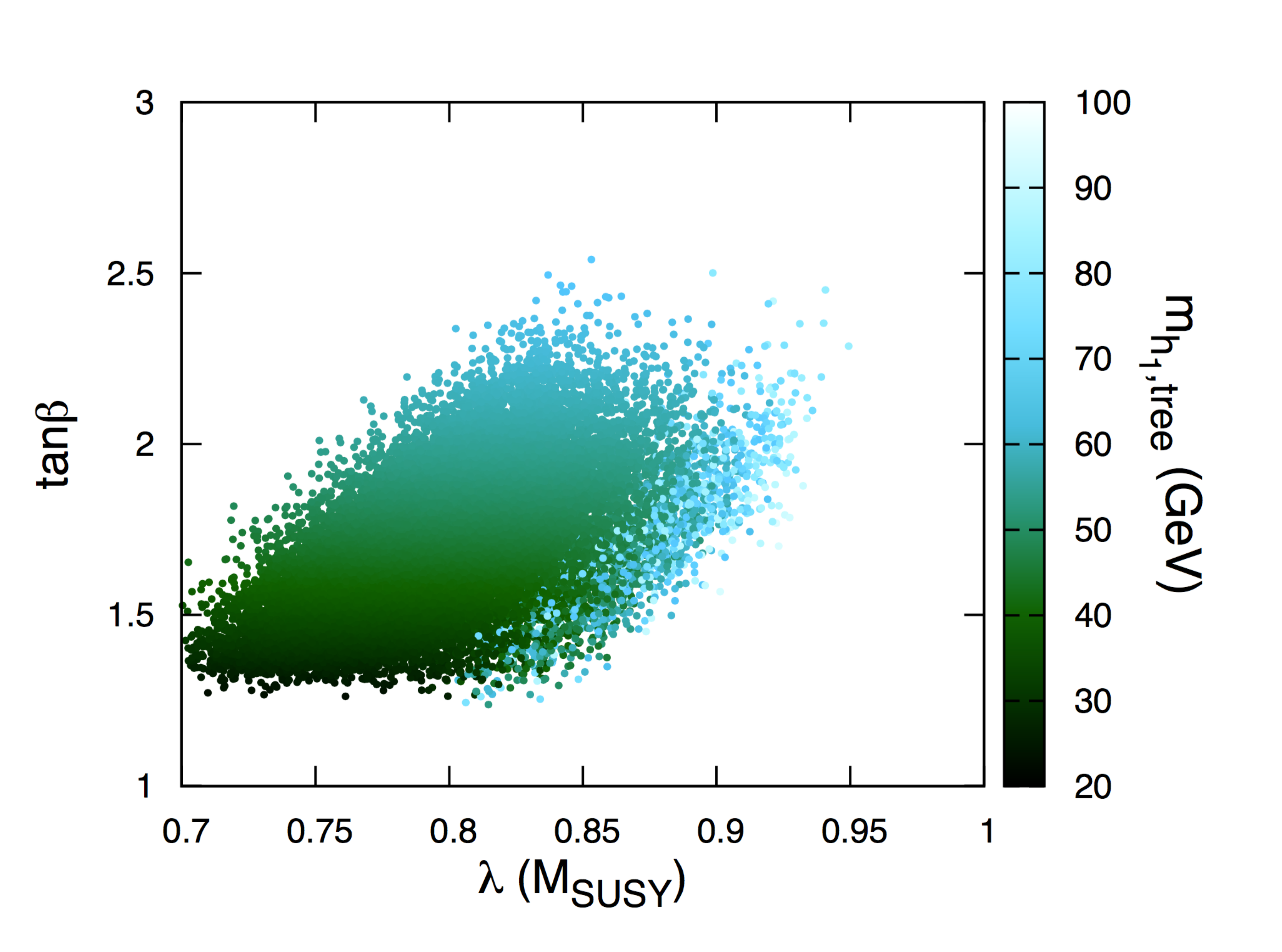}
}%
\caption{(a) The lightest Higgs mass value for the points within the $1\sigma$ theoretical uncertainty,
$122\gev\lesssim m_{h_1}\lesssim128\gev$, in the $(\lam(\msusy),\tanb)$ plane. 
(b) A zoom of the region of large $\lam$, where we plot the tree-level value of the Higgs mass in the $(\lam(\msusy),\tanb)$ plane.} 
\label{fig:param}
\end{figure}

In \reffig{fig:param}(a) we show the value of the lightest Higgs mass for the points within the $1\sigma$ theoretical uncertainty,
$122\gev\lesssim m_{h_1}\lesssim128\gev$, in the $(\lam(\msusy),\tanb)$ plane. 
One can easily recognize two different regions. 
The first is characterized by $\lam\approx 0.1-0.5$ and $\tanb\gtrsim 5$, and the scan seems to give there a Higgs mass slightly low, $m_{h_1}\lesssim124\gev$, 
albeit well within the adopted theoretical uncertainty. This ``small-$\lam$'' region shows solutions similar to the MSSM limit of the model~\cite{Jelinski:2014uba}. 
The Higgs mass decreases for lower values of \tanb, as a consequence of the 
rapid drop in the tree-level value, but it also decreases slowly for $\lambda>0.4-0.5$ when $\tanb\gtrsim 7$. 
For the majority of the points this is due to the trilinear term $|A_t|$ becoming smaller as $\lam$ 
increases,\footnote{Recall the RGEs for $A_t$: $dA_t/dt\sim\lam^2 A_\lam$~\cite{Ellwanger:2009dp}, where $A_\lam<0$, see, e.g., 
the second to last of \refeq{gmsbsoft} in Appendix~\ref{softs}.} 
as the scan cannot compensate this effect by raising \msusy\ because of our narrow range in $\Lambda$.

A second region, more interesting for the purposes of this paper, can be found instead at 
$\lam\gtrsim 0.7$ and $\tanb\lesssim 3$. As \reffig{fig:param}(a) shows, the Higgs mass there can comfortably reach 
the measured value and above. However, not all the points in this region are equivalent to each other, 
as an analysis of the tree-level Higgs mass can show. 

In \reffig{fig:param}(b) we present a zoomed-in detail of the ``large-$\lam$'' region, where we plot 
in the third dimension the tree-level value of the Higgs mass, whose explicit expression is presented in Appendix~\ref{app:higgs}.
Figure \ref{fig:param}(b) shows two distinct sets of points, 
characterized by different properties. For $\lam>0.8$ there are points characterized by tree-level masses
in the range $\sim70-90\gev$. These solutions are somewhat similar to 
models of $\lambda$SUSY~\cite{Barbieri:2006bg}, characterized by values of $\lam$ that can become non perturbative 
before reaching the GUT scale.\footnote{Since there are no direct couplings between the scalar $S$ and the messengers, and $S$ is also a gauge singlet, 
at one loop no additional terms arise in the $\beta_{\lambda}$ above $M_Y$. 
The gauge couplings $g_1$ and $g_2$ enter $\beta_{\lambda}$ and
in the presence of the messengers can get renormalized slightly more strongly. 
We have checked numerically that the effect is however very small and does not change the fact that $\lambda>0.8$ 
generally becomes nonperturbative at the GUT scale.}
The tree-level mass is enhanced with respect to the MSSM value, and as a consequence the needed radiative corrections 
are less substantial than in the MSSM for equivalent \tanb. 
Note, however, that we could not find a single region of the parameter space in which $m_{h_1,\textrm{tree}}^2>M_Z^2$, so that 
significant radiative corrections are \textit{always} necessary in this model to obtain the correct Higgs mass. 

Still, these points present somewhat lighter spectra than in the rest of the parameter space, although for the most part still too heavy
to be significantly probed in Run II at the LHC~\cite{ATL-PHYS-PUB-2014-010}. The lightest gluino mass found by the scan is around 2\tev, as can be seen 
in Table~\ref{tab:benchmarks}, where we present the parameters and spectral properties of a typical point (BP1).

\begin{table}[t]
   \centering\footnotesize
   \begin{tabular}{|c|c|c|c|c|} 
      \hline
      Benchmark & BP1 & BP2 & BP3 & BP4 \\
      \hline
      \hline
      Model parameters (at $M_Y$)&  &  &  & \\
       \hline
       $\Lambda$ & $2.98\times 10^5\gev$ & $5.79\times 10^5\gev$ & $5.92\times 10^5\gev$ & $9.15\times 10^5\gev$ \\
       \hline
       $M_Y$ & $2.49\times 10^{11}\gev$ & $5.99\times 10^{13}\gev$ & $9.56\times 10^{11}\gev$ & $2.48\times 10^{13}\gev$ \\
       \hline
       $h_3$ & $0.92$ & $0.65$ & $0.73$ & $0.23$ \\
       \hline
       \lam\ & $1.72$ & $1.69$ & $0.14$ & $0.35$  \\
       \hline
       $\xi_F$ & $2.14\times 10^{10}\gev^2$ & $2.03\times 10^{15}\gev^2$ & $1.23\times 10^{12}\gev^2$ & $3.70\times 10^{14}\gev^2$ \\
       \hline
       \hline
       Relevant for EWSB (at $M_{\textrm{SUSY}}$) &  &  &  & \\  
       \hline
       \hline
       \tanb\ & $1.70$ & $1.40$ & $7.52$ & $7.46$ \\
       \hline    
       $m_{H_d}^2$ & $4.33\times 10^7\gev^2$ & $4.39\times 10^7\gev^2$  & $8.95 \times 10^6\gev^2$  & $-8.13 \times 10^6\gev^2$ \\
	   \hline	
       $m_{H_u}^2$ & $-2.24\times 10^7\gev^2$ & $-3.19\times 10^7\gev^2$ & $-1.21\times 10^7\gev^2$ & $-1.51\times 10^7\gev^2$ \\
       \hline
       $\lambda$ & $0.82$ & $0.70$ & $0.13$ & $0.33$ \\
       \hline
       $\mu_{\textrm{eff}}$ & $7639\gev$ & $10746\gev$ & $3596\gev$ & $4048\gev$ \\
	   \hline
	   $A_\lam$ & $-1746\gev$ & $-1353\gev$ & $-7635\gev$ & $-1075\gev$ \\
       \hline    
       $m_S^2$ & $-4.90\times 10^7\gev^2$ & $-6.94\times 10^7\gev^2$ & $-1.71\times 10^6\gev^2$ & $-2.28\times 10^6\gev^2$ \\
	   \hline	
       $\mu'$ & $-1.69\times 10^{6}\gev$ & $-9.66\times 10^{10}\gev$ & $-4.53\times 10^{7}\gev$ & $-2.89\times 10^{10}\gev$  \\
       \hline
       $m_S'^2$ & $1.00\times 10^{12}\gev^2$ & $1.11\times 10^{17}\gev^2$ & $5.37\times 10^{13}\gev^2$ & $5.28\times 10^{16}\gev^2$  \\
       \hline
       $\xi_F$ & $1.58\times 10^{10}\gev^2$ & $1.49\times 10^{15}\gev^2$ & $1.23\times 10^{12}\gev^2$ & $3.58\times 10^{14}\gev^2$ \\
       \hline
       $\xi_S$ & $-9.17\times 10^{15}\gev^3$ & $-1.70\times 10^{21}\gev^3$ & $-1.44 \times 10^{18}\gev^3$  & $-6.55 \times 10^{20}\gev^3$ \\
       \hline
       \hline
       Max fine tuning & $\sim 10^4$ ($\lam,\xi_F$) & $\sim 10^6$ ($\lam$) & $\sim 10^4$ ($\xi_F$) & $\sim 10^4$ ($\xi_F$)\\
	  \hline
      \hline
      Spectrum &  &  &  & \\      				
    	  \hline
  	  \hline
  	  $m_{h_1,\textrm{tree}}$ & $74.1\gev$ & $27.7\gev$ & $84.6\gev$ & $84.1\gev$ \\
  	  $m_{h_1}$ & $123.5\gev$ & $123.0\gev$ & $122.6\gev$ & $122.3\gev$ \\
  	  $m_{h_2}, m_{a_1}$ & $5.86\times 10^4\gev$ & $2.23\times 10^7\gev$ & $7.11\times 10^4\gev$ & $5.91\times 10^6\gev$ \\
    	   $m_{\tilde{g}}$ & $2170\gev$ & $3959\gev$ & $4006\gev$ & $5970\gev$ \\
    	   $m_{\tilde{t}_1}$ & $2311\gev$ &  $3759\gev$ & $3436\gev$ & $5509\gev$ \\
    	   $m_{\chi_1^0}$ & $422\gev$ & $813\gev$ & $819\gev$ & $1266\gev$ \\
    	   $m_{\chi_1^{\pm}}$ & $832\gev$ & $1563\gev$ & $1550\gev$ & $2355\gev$ \\
    	   $m_{\tilde{\tau}_1}$ & $920\gev$ &  $2170\gev$ & $2391\gev$ & $261\gev$ \\
    	  \hline
   \end{tabular}
   \caption{The model fundamental parameters at $M_Y$, the corresponding parameter values at $\msusy$, 
   fine tuning, and spectra of four benchmark points discussed in Sec.~\ref{sec:pheno}. The parameters in parentheses give the maximal contribution to the fine-tuning measure.}
   \label{tab:benchmarks}
\end{table}


Several points in \reffig{fig:param}(b) are characterized by a relatively low value of the tree-level Higgs mass,
$m_{h_1,\textrm{tree}}^2\approx 20-60\gev$, despite sizeable values of \lam.
This can be understood from Eq.~\refeq{happr} in Appendix~\ref{app:higgs}, 
which shows that for low $\tanb$ the MSSM-like part of the tree-level Higgs mass is augmented in the PQ NMSSM by an extra term which 
is approximately $\delta m_{h_1,\textrm{tree}}^2\approx \lam^2 v^2\,\epsilon$, where $\epsilon=\xi_S/(\xi_F\mu')$ assumes 
in our model values in the range $\sim0-0.8$. Thus, the tadpole ratio significantly affects the value of the tree-level mass: if
$\xi_S\ll\xi_F\mu'$ the tree-level mass is approximately equal 
to its typical MSSM value, as is exactly the case for these points. 

For $\lambda\gtrsim 0.7$, these solutions can still quite comfortably reach $m_{h_1}\approx 125\gev$ 
thanks to very large $\lam$-dependent radiative corrections. A representative point of this region (BP2) can be found in Table~\ref{tab:benchmarks}. 
One can see that BP2 is characterized by a large messenger scale, $M_Y\approx 10^{14}\gev$, which is necessary to 
lift the heavy Higgs masses up and boost the Higgs-loops corrections to the lightest Higgs mass.
As can be expected, these points are marred by extremely high levels of fine tuning, because 
of the large sensitivity to $\lam$-driven radiative corrections. Thus, we will not discuss them further in what follows.  
 
The third benchmark point in Table~\ref{tab:benchmarks} (BP3) is 
representative of the small-$\lam$, almost MSSM-like region shown in \reffig{fig:param}(a) and discussed above.
With respect to the large-$\lam$ region (see BP1), the lightest points in the small-\lam\ region tend to present gluino and neutralino
masses approximately a factor 2 heavier than the lightest points with large \lam.
Note that the fine tuning of the small- and large-$\lam$ regions are very comparable. 

A discussion of the fine tuning of the model is presented in Sec.~\ref{sec:fine-tuning}.
The model is tuned at least to the level of $5\times 10^3$ -- $10^4$, but
it is interesting to note that, particularly for the points of the small-\lam\ region, which present equivalent 
MSSM solutions in the limit $\lam\rightarrow 0$, the fine tuning is not higher than it would be in the MSSM,
even if in the PQ NMSSM presented here 
we had to introduce additional large scale terms that enter the EWSB conditions: $\xi_F$, $\xi_S$, $\mu'$, and $m_{S}'^2$.
In fact, we will show in \refsec{sec:fine-tuning} that UV relations between these parameters allow for significant cancellations in the 
fine-tuning measure, so that this does not increase proportionally to the typical scale of the extra terms. 

The parameter $h_3$, which is responsible for the mixing between the messenger and matter sector, 
cannot span the entire range of Eq.~\refeq{h3range}: 
there is a gap in its allowed values, which was already observed for the MSSM in~\cite{Jelinski:2014uba}.
The reason can be easily understood by a quick analysis of Eqs.~\refeq{gmsbsoft} in Appendix~\ref{softs}.
The staus can become tachyonic when a cancellation takes place between terms dominated by the gauge couplings  
and those dominated by $h_3$. The position and range of this gap strongly depend 
on the values of \lam, and the gap tends to become smaller for larger \lam\ values.    

\begin{figure}[t]
\centering
\subfloat[]{%
\label{fig:a}%
\includegraphics[width=0.47\textwidth]{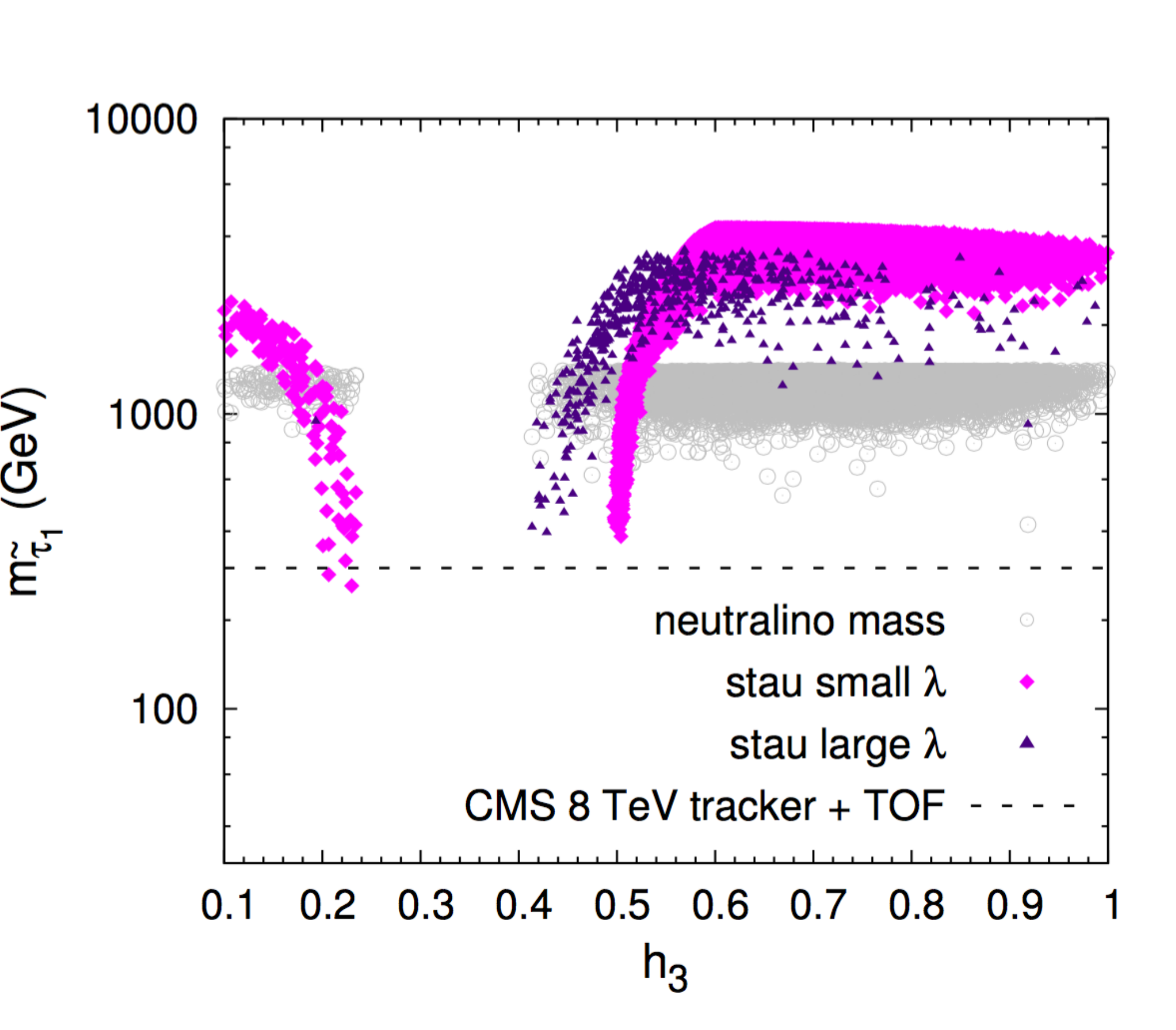}
}%
\subfloat[]{%
\label{fig:b}%
\includegraphics[width=0.47\textwidth]{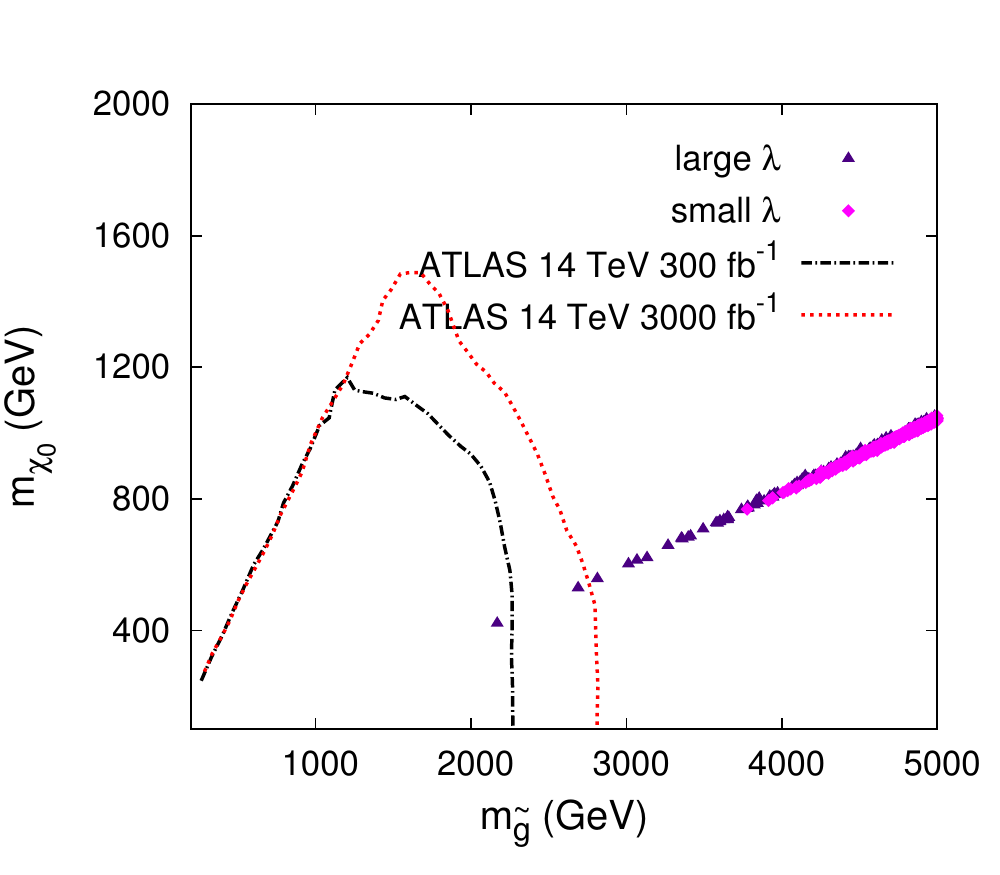}
}%
\caption{(a) The distributions of the lightest stau mass, $m_{\tilde{\tau}_1}$ as a function of $h_3$
for the points of the small-\lam\ region (magenta diamonds) and of the large-\lam\ region (indigo triangles). The neutralino mass is plotted in the background,
as gray circles. The dashed black line shows the 95\%~C.L. lower mass bound on heavy stable charged particles 
from the tracker + TOF analysis at CMS 8\tev~\cite{Chatrchyan:2013oca}. (b) The ATLAS projections for gluino 
searches at the LHC 14\tev~\cite{ATL-PHYS-PUB-2014-010} 
compared to the viable points of our model.
The dot-dashed black line shown the 95\%~C.L. expected exclusion reach with 300\invfb\ and the dotted red line the reach with 3000\invfb.}
\label{fig:spectr}
\end{figure}

In \reffig{fig:spectr}(a) we show the distribution of the stau mass as a function of $h_3$ 
for the small-\lam\ (magenta diamonds) and large-\lam\ (indigo triangles) regions. 
The stau mass can drop drastically as one approaches the critical values of $h_3$ giving rise to this cancellation,
to the point that it becomes the Next-to-lightest SUSY particle (the LSP is the gravitino), as a 
comparison with the neutralino mass, plotted here with gray circles, shows. 

This brings about the interesting possibility of testing selected $h_3$ regions by searches for heavy stable charged particles.
We plot in \reffig{fig:spectr}(a) with a black dashed line the 95\%~C.L. 
lower bound on the mass of the stable stau from an analysis of long time-of-flight (TOF) 
to the outer muon system and anomalously high (or low) energy deposition in the inner tracker at CMS, with 
8\tev\ data~\cite{Chatrchyan:2013oca}. The search excludes cross sections of the order of $1.5$~fb or above for pair production
of $\sim300\gev$ stable staus, which is in agreement with what one obtains in this model with typical spectra and a stau
NLSP with a mass of 300\gev.  
We thus expect that the corresponding search in Run II will start biting into specific regions 
of our parameter space: $h_3\approx 0.2$ or $0.5$ for $\lam<0.4$; $h_3\approx 0.4$ for $\lam\approx 0.85-0.9$. 
We show a representative point with stau NLSP (BP4) in Table~\ref{tab:benchmarks}. 

We conclude this section by comparing in \reffig{fig:spectr}(b) the projected~\cite{ATL-PHYS-PUB-2014-010} exclusion 
reach for gluino masses at ATLAS 14\tev\ with our model's predictions. 
Again, points of the large-\lam\ region are shown as indigo triangles and points of small \lam\ as magenta diamonds.
The 95\%~C.L. expected reach with 300\invfb\ of integrated luminosity in searches with jets + missing $E_T$ is shown as 
a dot-dashed black line.
As is well known, in GMSB models specific LHC phenomenology is strongly affected by the effective coupling of  
the neutralino (which is predominantly bino-like in our model) and the gravitino. 
However, gluino searches at 8\tev\ have shown that the 
mass bounds do not significantly depend on whether the neutralino 
is long-lived enough to escape the detector~\cite{Aad:2014wea,Khachatryan:2015vra}, 
or it decays promptly producing photons~\cite{Aad:2015hea}.
Thus, for the purpose of this paper we accept the projected reach of~\cite{ATL-PHYS-PUB-2014-010} as a reasonable estimate.

One can see in \reffig{fig:spectr}(b), that the points of the large-\lam\ region might begin to be probed at the very end of the LHC Run II, 
particularly if the High-Luminosity LHC~\cite{ATLAS}, whose reach~\cite{ATL-PHYS-PUB-2014-010} is shown as a dotted red line, is approved.  
However, \reffig{fig:spectr}(b) more realistically shows that the structure devised here 
yields spectra too heavy to be significantly tested at the LHC, and a high-energy collider will be necessary. 
It might be interesting to study the reach of a 100\tev\ collider for this model, and we leave 
this for future work.

\section{Fine tuning\label{sec:fine-tuning}}

We dedicate this section to the calculation of the fine tuning of the model.
As was mentioned in Sec.~\ref{sec:pheno}, it turns out that this is smaller than what one could expect 
from naive estimates based on a rough order-of-magnitude analysis.
We take the maximum of the fine tuning due to each of the fundamental parameters of Eq.~\refeq{pars}, 
which we calculate according to the Barbieri-Giudice measure~\cite{Ellis:1986yg,Barbieri:1987fn}.

By indicating the parameters of Eq.~\refeq{pars} collectively as $p_i$,
one must calculate the max of the
\begin{equation}
\Delta_{p_i}=\frac{\partial \log M_Z^2}{\partial \log p_i}=\frac{p_i}{M_Z^2}\frac{\partial M_Z^2}{\partial p_i}\,\,\,\textrm{ (order of magnitude only)}\,.
\end{equation}

Given the three EWSB conditions that determine the values of the Higgs 
vevs,\footnote{The explicit forms of $\textrm{EWSB}_1$, $\textrm{EWSB}_2$, and $\textrm{EWSB}_3$ can be found in Eqs.~\refeq{ext}-\refeq{extrema} of 
Appendix~\ref{estimates}.}
\begin{equation} 
\textrm{EWSB}_j(M_Z^2,\tanb,s,\mu',\xi_S,m_{S}'^2...; p_i)=0\,\,\,\,\textrm{for $j=1,2,3$}\,,
\end{equation}
one can apply the chain rule,
\begin{equation}
\frac{\partial P_k}{\partial p_i}=\left[\frac{\partial (\textrm{EWSB}_j)}{\partial P_k}\right]^{-1}\frac{\partial (\textrm{EWSB}_j)}{\partial p_i}\,,\label{chain}
\end{equation}  
where the $3\times 3$ matrix in the $jk$ indices is built out of the partial derivatives with respect to $P_k=\{M_Z^2,\tanb,s\}$\,, 
and one is interested in the numerical value of the first row in the system \refeq{chain}, $\partial M_Z^2/\partial p_i$\,.

Despite the presence of large contributions to EWSB from the tadpoles and other terms, 
a numerical calculation shows that the partial derivatives are sensitive to cancellations. 
To see this, we parametrize two large parameters, $\xi_S$ and $m_S'^2$\,,  in the following way:
\beq\label{rel}
\xi_S=-a \Lambda \xi_F,\quad m_S'^2 = -b \Lambda \mu'\,,
\eeq
where $a$ and $b$ are arbitrary real numbers, and we proceed to the calculation of $\Delta_{\xi_F}$ and $\Delta_{M_Y}$ for changing $a$ and $b$\,. 
$M_Y$, which directly enters the definition of $\mu'$, and $\xi_F$ are the largest remaining free parameters in the theory.

\begin{figure}[t]
\begin{center}
\includegraphics[width=0.45\textwidth]{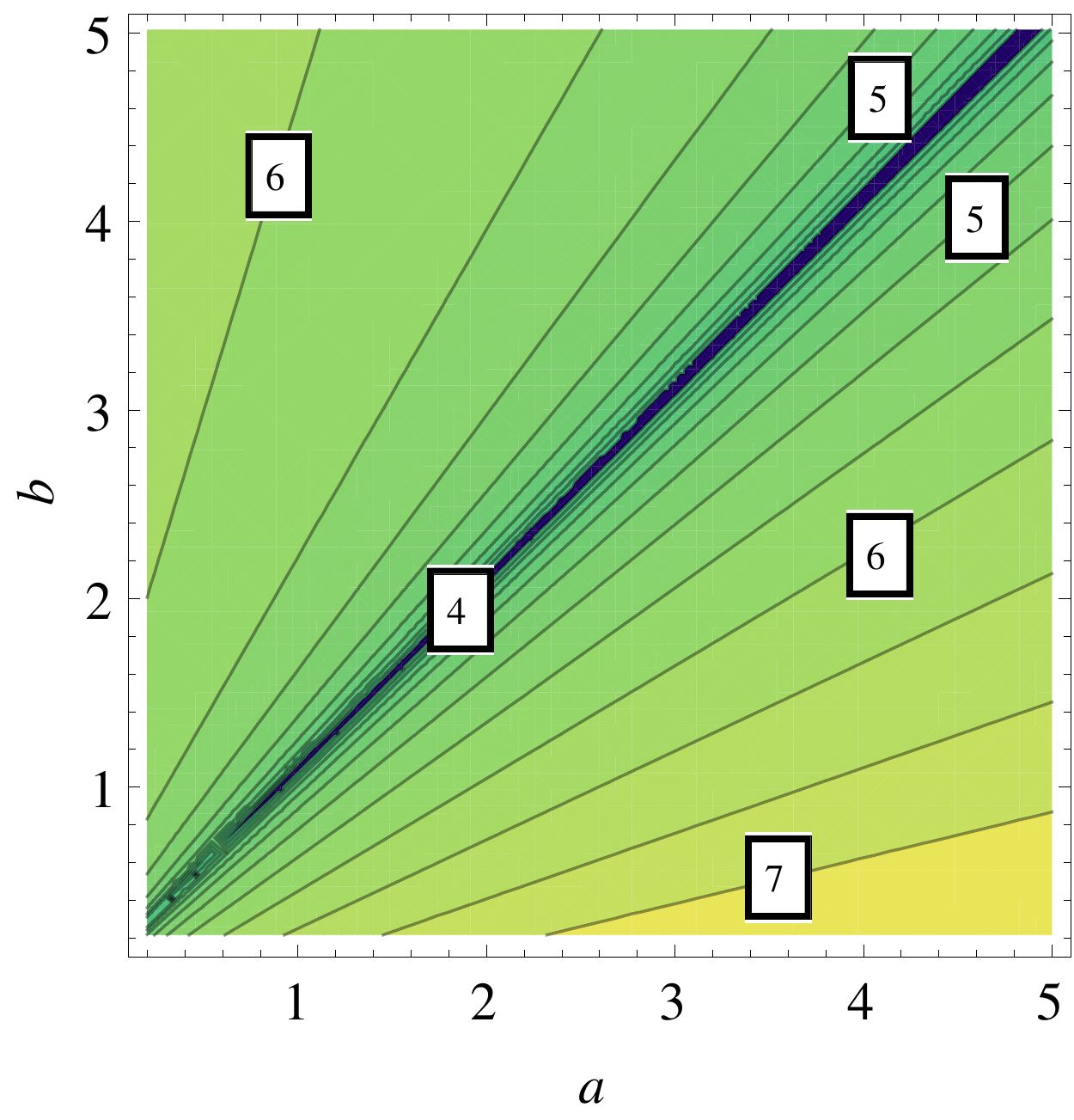}
\caption{A plot of $\log_{10} \Delta_{\xi_F}$ for BP1 of Table~\ref{tab:benchmarks} as a function of the arbitrary real coefficient $a,b$\,.}
\label{fig:finetune}
\end{center}
\end{figure}

We show in Fig.~\ref{fig:finetune} a plot of $\log_{10} \Delta_{\xi_F}$ in the $(a,b)$ plane for BP1. 
A plot of $\log_{10} \Delta_{M_Y}$ shows similar behavior and we do not present it here.
One can clearly see that the fine tuning is reduced by more than two orders of magnitude 
when $a\approx b$. 

One can see from Eqs.~(\ref{B8}) and (\ref{B9}) of Appendix~\ref{app:eff} that in our model $b=2$ is a consequence of breaking SUSY via the spurion. 
As a consequence, according to Fig.~\ref{fig:finetune}, one needs $a\approx 2$ to lower the fine tuning to $\sim10^4$. 
As is explained in Appendix~\ref{app:eff} after Eq.~(\ref{instantons}), we make here a reasonable assumption, that there exist selection rules in the stringy UV completion 
that allow one to generate only the instanton contributions that lead to $a=2$.
Note that, as one can see in Eqs.~(\ref{fulleqs}) and following paragraph, 
the terms depending on $\lambda_{\mu}$ are much smaller than the instanton contribution to $\xi_S$, so that $\lambda_{\mu}\neq 0$ will not 
cause significant deviations from $a=2$. 

The cancellations described here have their origin in the specific form of the tree-level EWSB conditions. 
They lead to the fine-tuning levels of the benchmark points in Table~\ref{tab:benchmarks}, and in general
similar values are obtained over the whole viable parameter space. 
Thus, as was anticipated in \refsec{sec:intro}, our model is partially protected 
from the tadpole problem, provided an appropriate relation of the kind of \refeq{rel} is given by the UV completion. 

\section{Summary and conclusions}\label{sec:summary}

In this paper, motivated by the desire of maintaining a Yukawa flavor structure in agreement with 
UV completions based on F-theory, we have used flavored gauge mediation to construct a specific version of the Peccei-Quinn NMSSM.
Considerations concerning the stringy UV completion lead to a very predictive version of the model, which depends on a few free parameters.

We have performed a thorough numerical study of the parameter space of the model and confronted it with the phenomenological constraints. 
Our findings are supported by analytical arguments. 
We showed that the PQ NMSSM exhibits unusual properties, specifically it requires relatively large nonstandard terms:
tadpole coefficients $\xi_F$, $\xi_S$, and quadratic terms depending on $\mu'$ and $m_S'^2$.  
In spite of this, we showed that the fine tuning is not greater than in the MSSM thanks to a special relation between the
ratios $\xi_S/\xi_F$ and $m_S'^2/\mu'$, which originates in the specific UV completion considered here, and which results 
in a cancellation in the fine-tuning measure.
Thus, we showed that this version of the the PQ NMSSM is partially immune from the mild tadpole problem typical of many general versions of the NMSSM. 

We found that the model can easily accommodate the Higgs mass at 125\gev,
but the spectra can possibly be in reach of the LHC only in a region 
of the parameter space characterized by $\lam>0.8$ which, as is well known, 
implies that the renormalized value of \lam\ at the GUT scale might incur a Landau pole.
In general, we were not able to find a single parameter space region over which the tree-level value 
of the lightest Higgs mass is larger than the $Z$ boson mass, although it can be enhanced with respect to its MSSM counterpart in 
regions of large \lam\ and small \tanb.
In all cases, however, $m_{h_1}\approx125\gev$ requires substantial radiative correction and
a certain level of fine tuning, of the order of $10^4$, cannot be avoided.

In spite of these interesting formal properties, we found that the phenomenology of the model is quite standard, 
with typical spectra being too heavy
to be significantly probed at the LHC. However, small regions of the parameter space 
exist featuring gluino masses around 2\tev, for a bino-like neutralino mass of $\sim400\gev$. 
These might begin to be probed at the end of Run II in gluino searches.
For particular choices of the parameter $h_3$, which in our model determines the mixing between the messengers and matter sector, 
the NLSP is a stau with a mass $\gtrsim 300\gev$, and a cross section right in the ballpark of the range 
currently probed by searches for heavy stable charged particles at CMS, 
which will be able to further constrain part of the parameter space with
13 and 14\tev\ data. 


This work can be extended in several directions. 
An open question remains concerning the dynamics of the spurion $X$, which should result from a UV completion of the model
(supergravity or a string compactification with fluxes). 
It is of course possible that most of the necessary ingredients are contained in Appendix~\ref{app:eff}, with the exception of some self-interaction terms in $X$. 
In that case, including spurion quanta into the analysis should not pose a difficult task.

Another possibility would be to enlarge the size of the corrections given by exchange of the \upq\ gauge bosons by increasing the strength of 
the $g_{PQ}$ coupling. This would amend the soft terms given in Appendix~\ref{softs} by possibly substantial corrections.
Finally, one could lower the messenger mass, which would result in a modification of the soft masses by one-loop terms.

The list of modifications is of course much longer, and we leave this extended discussion to a future publication.

\bigskip
\noindent \textbf{Acknowledgments}
\medskip

\noindent We would like to thank Tomasz Jelinski and Andrew Williams for helpful discussions and inputs. 
K.K. is supported in part by the EU and MSHE Grant No. POIG.02.03.00-00-013/09.
K.K. and E.M.S. are funded in part by the Welcome Programme of the Foundation for Polish Science. 
The use of the CIS computer cluster at the National Centre for Nuclear Research is gratefully acknowledged. 
\bigskip


\newpage

\appendix

\section{Soft terms}\label{softs}

At the leading order, all soft masses and trilinear couplings are generated at the messenger scale $M_Y=\vev X$. 
While deriving the explicit formulas, we will make some simplifying assumptions. 

First of all recall that Eq.~\refeq{leadY} is defined at \lgut. Evolution to $M_Y$ leads to further mixing between matter and messengers.
However, we have explicitly checked that this has a very mild influence on the structure and values of the SM Yukawa couplings and
$h_3$ at $M_Y$\,.
Secondly, we take into account only the leading superpotential contributions, Eq.~\refeq{leadY}, 
thus disregarding all other contributions from Eq.~\refeq{mainY}, which are usually much smaller. 
In particular, we neglect the terms proportional to the bottom and tau Yukawa couplings.

In the end, all the soft terms depend only on $\lsusy=F_X/\vev X$, $\la$, $h_3=\widehat{y}_{34}$, $y_t=y^u_{33}$, and the gauge coupling constants, 
while the dependence on the scale $M_Y$ appears through the RG evolution. 

We adopt here the general formulas derived in~\cite{Chacko:2001km,Evans:2013kxa} in the presence of MSSM-messenger 
superpotential interactions, which provide a good approximation for the case at hand.
The resulting soft terms are thus
\beqa\label{sm-q}
(m_{\widetilde Q})^2_{ij}&=&\frac{\lsusy^2}{3840\pi^4}\left\{\d_{ij}8\pi^2\left(\alpha_1^2+45\alpha_2^2+80\alpha_3^2\right)
+\d_{i3}\d_{j3}|h_3|^2 \left[105 |h_3|^2-4\pi\left(7\alpha_1+45\alpha_2+80\alpha_3\right)\right]\right\},\non
(m_{\widetilde U})^2_{ij}&=&\frac{\lsusy^2}{30\pi^4}\left[\d_{ij}\pi^2\left(\alpha_1^2+5\alpha_3^2\right)-
\d_{i3}\d_{j3}\frac{15}{64}|y_t|^2|h_3|^2\right],
\non
(m_{\widetilde D})^2_{ij}&=&\frac{\lsusy^2}{120\pi^4}\left[\d_{ij}\pi^2\left(\alpha_1^2+20\alpha_3^2\right)\right],
\non
(m_{\widetilde L})^2_{ij}&=&\frac{3\lsusy ^2}{160\pi^4}\left[\d_{ij}\pi^2\left(\a_1^2+5 \a_2^2\right)\right],\non
(m_{\widetilde E})^2_{ij}&=&\frac{\lsusy^2}{640\pi^4}\left\{\d_{ij}48\pi^2\alpha_1^2+\d_{i3}\d_{j3}|h_3|^2\left[35|h_3|^2-12\pi\left(3\alpha_1+5\alpha_2\right)\right]\right\},\non
m_{H_{u}}^{2}&=&\frac{3\lsusy^{2}}{160\pi^{4}}\left[\pi^{2}(\alpha_{1}^{2}+5\alpha_{2}^{2})-\frac{5}{8}|h_{3}|^{2}|y_t|^{2}-\frac{5}{6}\lam^2 |h_3|^2\right],\non
m_{H_{d}}^{2}&=&\frac{3\lsusy^{2}}{160\pi^{4}}\left\{\pi^{2}(\alpha_{1}^{2}+5\alpha_{2}^{2})+\frac{1}{24}|h_{3}|^{2} \left[ 140|h_{3}|^{2}+15
|y_t|^{2}-16\pi(4\alpha_{1}+15\alpha_{2}+20\alpha_{3})\right]\right\}\,,\non
A_t&=&-\frac{\lsusy}{16\pi^{2}}|h_{3}|^2\,,\non
A_\lam&=&-\frac{4 |h_3|^2}{16\pi^2}\Lambda\,,\non
m_S^2&=&-\frac{\lam^2 |h_3|^2}{32\pi^4}\Lambda^2\,,\label{gmsbsoft}
\eeqa
According to the SLHA2 convention~\cite{Allanach:2008qq} that we employ here, 
$A_{t}$ is related to the ``up'' trilinear coupling $(T_{u})_{33}$ as $A_t=(T_{u})_{33}/y_t$\,.\footnote{Recall 
that intergenerational up trilinear couplings enter the scalar potential, 
$V$, as $V\supset-H_{u}\widetilde{Q}_{i}(T_{u})_{ij}(\widetilde{u}_{R}^{*})_{j}$\,.}
Note also that we choose $\Lambda>0$\,, 
so that both $A_\la$ and $m^2_S$ are bound to assume exclusively negative values.

At the leading order (one loop) gaugino masses at $M_Y$ are directly related to the gauge coupling constants and given by
\beq
M_{i}=\frac{\lsusy\a_{i}}{4\pi}\quad \textrm{for}\; i=1,2,3.
\eeq

As was shown in the numerical analysis of \refsec{sec:pheno}, 
some of the soft masses become negative for some $h_3$ values. 
As an example, one can derive from Eqs.~\refeq{gmsbsoft} some naive bounds, 
where for simplicity all $\a$'s are set equal to $1/20$ at $M_Y$, $y_t\approx0.6$, and $\tanb<10$\,:  
\beq
h_3<\frac1{12}\quad\mbox{or}\quad \frac14<h_3<1.3\,.
\eeq
However, the above values are just indicative. The full numerical bounds are shown in \refsec{sec:pheno}.

\section{Sources of effective interactions below the messenger scale\label{app:eff}}

We will discuss here corrections to the dynamics of $S$ due to various processes that might take place in
a UV completion of our PQ NMMSM.\footnote{For a different approach see, e.g., the analysis of Ref.~\cite{Im:2014sla}.} 
We base our arguments on certain F-theoretic constructions. As explained, e.g., in Refs.~\cite{Beasley:2008dc,Beasley:2008kw}, 
chiral matter originates from two D7-branes intersecting along a 2-dimensional Riemann surface; 
matter interactions come from the intersection of three D7-branes at one point, which results in the intersection
of three matter Riemann surfaces at this point. These intersections are generic, in the sense that they naturally emerge in the 
geometry of the string compactification. Thus, at the compactification 
scale the only natural matter couplings are cubic, and for this reason we set the mass term $X S$ to zero in \refsec{sec:gmsbpq}. 
The same argument also suggests that all natural couplings in the superpotential should be of the order of 1.
The simple picture just described is generally amended in the presence of family effects~\cite{Cecotti:2009zf}, instantons, which we discuss below, and possibly other 
effects that we do not discuss here and are difficult to estimate without a detailed knowledge of the full F-theory model.

We include in the analysis all important terms in the spurion $X$, whose vev gives a mass to the messengers and breaks SUSY.
Below the messenger scale, $M_Y\equiv\langle X\rangle$, the spurion will be
a nondynamical field, i.e., we set $X\rightarrow \vev X+\theta^2 F_X$\,.
$F_X$ will be the primary source of SUSY breaking, which means that $F$-terms of other fields must be much smaller
($\xi_F\ll F_X$ in what follows). Note that $\vev X$ breaks also \upq. 
This means that at low energies there appear new effective \upq\ breaking interaction terms.

The contributions that might be relevant below $M_Y$ can have various sources. For the low-energy dynamics
the most important will be terms with powers of the spurion superfield $X$, because $X$ receives a large vev. 
As usual, operators of dimension 5 and higher will be suppressed by a large mass scale of the order of \lgut\ or
even $\mpl$.

\paragraph{Giudice-Masiero term.}
Besides the superpotential term (\ref{lam}),
the only cubic couplings of $S,\,X$ allowed by \upq\ is in the 
K\"{a}hler potential:
\beq\label{d-cubic}
\d K=\la_\mu \frac{X^{\dag}}{\lgut}H_u H_d+\textrm{h. c.}\,,
\eeq
where $\la_\mu$ is a complex constant difficult to estimate~\cite{Heckman:2008qt}.

The K\"{a}hler potential $\delta K$ can generate 
an additional $\mu$ term in the superpotential when the spurion $X$ gets its vev, like in the Giudice-Masiero~\cite{Giudice:1988yz} mechanism:
$\mu_{\textrm{GM}}=\lam_\mu F^{\ast}_X/\lgut$\,.
However, one can either choose $\lam_\mu=0$, or perform a redefinition of the field, 
$S\rightarrow S-\mu_{\textrm{GM}}/\lam$, which makes $\mu_{\textrm{GM}}$ disappear from the superpotential, 
and at the same time generates corrections to the tadpoles, $\delta \xi_F$ and $\delta \xi_S$, and a $B\mu$ term, $m_3^3$, of the form:
\beqa
\delta\xi_F&=&-\frac{\lam_{\mu}}{\lam}\mu'\Lambda x\,,\non
\delta\xi_S&=&-\frac{\lam_{\mu}}{\lam}m_S'^2\Lambda x-\frac{\lam_{\mu}^{\ast}}{\lam^{\ast}}m_S^2\Lambda x\,,\non
m_3^2&=&-\lam_{\mu}A_{\lam}\Lambda x\,,
\eeqa
where $\mu'$, $m_S'^2$, $m_S^2$, and $A_\lam$ are given in Eqs.~(\ref{eff_model}) and (\ref{sm-q}).

Note that in both cases the only origin of the $\mu$-term remains the vev of the singlet, $s=\vev S$, so that
$\mu_{\textrm{eff}}=\lam s$.


\paragraph{Instantons.}
In stringy models there are several types of instantons~\cite{Witten:1996bn,Douglas:1996sw}. 
An instanton action, $S_{\mathcal{I}}$, depends on various moduli fields, which generally belong to two main categories, 
``twisted'' and ``untwisted''. 
Twisted instantons can transform by a shift under some abelian gauge symmetry, which in our case means that 
our \upq\ effectively gives a charge to $e^{-S_{\mathcal{I}}}$. 
When the moduli are set to some fixed value, $e^{-S_{\mathcal{I}}}$ becomes a numerical factor, which we denote here as $I_{q,n}$.
In this case $q$ is the \upq\ charge carried by the instanton and $n$ indicates how many fields are associated with it, as shown below. 
Some of the instanton contributions are expected to vanish, while the ones that are present are expected to 
be strongly suppressed, $I_{q,n}\ll 1$. 

Instanton effects allow one to generate new terms in the superpotential (corrections to the K\"{a}hler potential are negligible):
\beq\label{inst-gen} 
\d W_{\mathcal{I}}=\mpl^2 I_{-1,1}\,S+\mpl( I_{0,2}\,XS +I_{-2,2}\,S^2)+
I_{1,3} X^2 S+I_{-1,3}X S^2+I_{-3,3}S^3+...
\eeq
where $\mpl$ is the reduced Planck mass and the ellipsis denotes higher powers of the fields, and we have here suppressed all terms depending on $X$ solely. 

As one can see, $\k\approx 0$, whereas other possible contributions might be important 
due to the large values of $\mpl$, $\vev X$, and $F_X$.
The surviving terms are
\beqa\label{d-inst}
\d W_{\mathcal{I}}&\supset &\left( \mpl^2 I_{-1,1}+\mpl \vev X I_{0,2}+I_{1,3} \vev X^2\right) S+\left(\mpl I_{-2,2}+I_{-1,3}\vev X\right) S^2\non
 &=&\mpl^2 \left( I_{-1,1}+\eta I_{0,2}+\eta^2 I_{1,3} \right) S +\mpl\left( I_{-2,2}+\eta I_{-1,3}\right) S^2\,,\label{instantons}\\
\d L_{\mathcal{I}}&=& F_X\mpl\left( I_{0,2} +2\eta  I_{1,3}  \right) S+I_{-1,3}F_X S^2\,,
\eeqa
where $\eta=\vev X/\mpl \ll 1$. 

From Eq.~\refeq{instantons} follows that if all $I_{q,n}$ are approximately of the same order of magnitude 
the most important contributions are those characterized by the lowest \upq\ charges, $q$. 
However, it is not uncommon for the dynamics of the UV stringy theory to generate selection rules that can forbid 
certain terms. If, for example, $I_{q\leq 0,n}=0$ due to the UV dynamics, one is left only with
\beqa\label{d-inst1}
\d W_{\mathcal{I}}&=& \vev X^2 I_{1,3}  S\,, 
\non
\d L_{\mathcal{I}}&=& 2 F_X \vev X  I_{1,3}   S\,.
\eeqa
In our study we always make this assumption, as it will be beneficial to reducing the fine tuning of the model.

\paragraph{Exchange of a heavy chiral \boldmath$U(1)_{\textrm{PQ}}$ neutral superfield.}
In general, string theory constructions imply the presence of some \upq\ heavy chiral neutral superfields 
at the GUT scale that can be integrated out. 

The simplest superpotential coupling with one heavy chiral state
$\Psi$ reads: $W\supset \la_\Psi XS\Psi+\frac12 M_\Psi \Psi^2$. 
Integrating out $\Psi$ gives
\beqa\label{d-psi}
\d W_{\Psi}&=&-\frac{\la_\Psi^2}{2M_\Psi} (XS)^2,\quad \d K_{\Psi}=\frac{|\la_\Psi|^2}{M_\Psi^2} (XS)^{\dag} (XS)\,,
\eeqa
which after SUSY breaking yields
\begin{equation}
\d W_{\Psi}=-\frac{\la_\Psi^2}{2M_\Psi} \vev X^2 S^2=\frac{1}{2}\mu'S^2\,,\label{B8}
\end{equation}
and
\begin{equation}
\d L_{W,\Psi}=\frac{\la_\Psi^2}{M_\Psi}F_X \vev X S^2+\textrm{h. c.}=-\frac{1}{2}m_S'^2 S^2+\textrm{h. c.}\label{B9}
\end{equation}
from the superpotential and 
\begin{equation}
\d L_{K,\Psi}=\frac{|\la_\Psi|^2}{M_\Psi^2}  |F_X|^2 |S|^2=-\delta m_S^2\, |S|^2
\end{equation}
from the K\"{a}hler potential (see Sec.~\ref{sec:eff}).

As was explained in Sec.~\ref{sec:eff}, we expect $M_\Psi\approx\lgut$ and $\la_\Psi\approx 1$ if nonvanishing.

\paragraph{Exchange of a \boldmath$\upq$ heavy gauge boson.}
Exchange of a heavy gauge boson $V$ associated with the breaking of \upq\ can also contribute to the 
scalar Lagrangian:
\beq
\d K_{V}=-\frac{g_{PQ}^2 \qpq_X\, \qpq_j}{M_V^2} (X^{\dag} X)(\Phi_j^{\dag}\Phi_j)\,\,\,\Rightarrow\,\,\,
\d L_V= -\frac{g_{PQ}^2 \qpq_X\, \qpq_j}{M_V^2} |F_X|^2 \Phi_j^{\dag}\Phi_j\,\,\,\,\textrm{ for all $j$}\,,\label{vecexch}
\eeq
where $\Phi_j$ are the \upq\ charged fields, $q_j$ are their charges, $g_{PQ}$ is a generic coupling constant, 
and $M_V\approx\lgut$. Equation (\ref{vecexch}) implies that all the soft masses squared 
$\widetilde{m}_j^2$ will receive an additional correction 
of the order of $g_{PQ}^2 q_X q_j \Lambda^2 x^2$\,.

\paragraph{Other contributions.}
There are more terms preserving \upq\ which might renormalize the K\"{a}hler potential as, for example, $\d K=\frac{1}{M^2} X^{\dag} XXS+\textrm{c. c.}$ 
With $M\approx\lgut$ they are negligible, so that we will not try to pinpoint their sources here.

\paragraph{Leading supergravity corrections.}
Supergravity corrections~\cite{Binetruy:2004hh} for $W=W_0+F_X X+ \xi_F S+\frac12\mu' S^2$ (with $\xi_F\ll F_X$)
and for the canonical K\"{a}hler potential subject to the condition of vanishing cosmological constant read: 
\begin{equation}
\d V(S)=\left(-2\frac{F_X^* \xi_F}{\sqrt3 \mpl}\ S-\frac{F_X^*\mu'}{\sqrt3\mpl}\ S^2+\textrm{c.c.}\right)+
4\frac{|F_X|^2}{3\mpl^2}\ |S|^2+...
\end{equation}
They are also negligible.

\bigskip

In summary, when expressed in terms of $\Lambda=F_X/M_Y$ and $x=M_Y/\lgut$, 
the full corrections to the superpotential and soft SUSY breaking terms are the following:
\beqa
\delta m_i^2&=&g_{PQ}^2\, q_X q_i |\lsusy|^2 x^2\non
\delta m_S^2&=&-|\la_\Psi|^2|\lsusy|^2 x^2+g_{PQ}^2\, q_X q_S |\lsusy|^2 x^2 \non
\mu'&=&-\la_\Psi^2 \lgut\ x^2,\non
m_S'^{2}&=& 2\la_\Psi^2\lsusy\lgut\ x^2=-2\mu' \lsusy\,,\non
m_3^2&=&-\lam_{\mu}A_{\lam}\Lambda x\,,\non
\xi_F&=&I_{1,3} \lgut^2 x^2+\frac{\lam_{\mu}\lam_{\Psi}^2}{\lam}\lgut\Lambda\,x^3,\non
\xi_S&=&-2I_{1,3}\lsusy \lgut^2 x^2 -\frac{2\lam_{\mu}\lam_{\Psi}^2}{\lam}\Lambda^2 x^3\lgut+\frac{\lam_{\mu}^{\ast}\lam_{\Psi}^2}{\lam^{\ast}}\Lambda^3 x^3-
\frac{\lam_{\mu}^{\ast}}{\lam^{\ast}}m_S^2\Lambda x\,,\label{fulleqs}
\eeqa
where in the last equation of (\ref{fulleqs}) $m_S^2$ is the term calculated in Appendix~\ref{softs}.

Since typical values of $\Lambda$ imply that $\Lambda x\ll \lgut$, 
in our study we assume $\lam_{\mu}=0$ without loss of generality. Moreover, $\Lambda^2 x^2\ll \msusy^2$
so that we can neglect the $g_{PQ}$ corrections to the soft masses.
Note that when $\lam_{\mu}=0$ and $g_{PQ}=0$\,, Eqs.~(\ref{fulleqs}) reduce to the form given in Eqs.~(\ref{eff_model}).

\section{Estimates of constants}
\label{estimates}

In this appendix we estimate the order of magnitude of different coefficients appearing in the Lagrangian. 
For simplicity in what follows all coefficients are assumed to be real.
We consider constraints that come from the tree-level vacuum equations of motion (e.o.m.), and those that follow from the 
stability of the physical vacuum. 

Let us recall that $\k=0$, $A_\k=0$, and we are interested in small \tanb\ values.
Moreover, we assume that all the GMSB soft masses given in Appendix~\ref{softs} are of the order of $\msusy\ll\Lambda$. 
We also assume that no accidental cancellation takes place between different terms. 
We only consider terms in leading powers of $v$, and do not explicitly differentiate between constants of the order of a few.

The vacuum e.o.m. are given by:
\beqa
\sin 2\beta-\frac{2(\la s\,A_\la +\widehat{m}_3^2)}{2\la^2s^2+\la^2\,v^2+m_{H_d}^2+m_{H_u}^2}
&=&0\,,\label{ext} \\
\left(\la^2-g^2\right)\,v^2 +\tan 2\beta\left(-m_{H_d}^2+m_{H_u}^2\right)
-\frac{2}{\tan 2\beta}\left(\la s\, A_\la +\widehat{m}_3^2\right) &=& 0\;, \\
s\left(m_{S}^2 +m_{S}'^2 +\mu'^2 
+\la^2\,v^2\right)+\xi_S+\xi_F \mu' -\la v^2\,\frac{\sin 2\beta}{2}\left(A_\la+\mu'\right)&=& 0\,,
\label{extrema}
\eeqa
where $g^2=(g_1^2+g_2^2)/2$ and $\widehat{m}_3^2= m_3^2 + \lambda(\mu' s + \xi_F) $.
Short inspection of the e.o.m. reveals that for $\tanb$ equals a few, $s$ is of the order of the soft SUSY masses, $s\sim \msusy$. 
Thus, since $\sin 2\beta\sim 1$ one also finds $\widehat{m}_3^2\sim \mu' s + \xi_F\lesssim \msusy[2]$. 

On the other hand, the requirement of tree-level stability in the scalar sector yields
\footnote{We analyze here the diagonal minors of the scalar mass matrix, which can be explicitly found, e.g., in Ref.~\cite{Ellwanger:2009dp}.}
\beqa
\left[-2s \la +(A_\la+\mu')\sin 2\beta\right]^2<-\frac{\left(\xi_S+\xi_F\mu'\right)}{s}\,,
\eeqa
which implies
\begin{equation} 
-(\xi_S+\xi_F\mu')/s\gtrsim\max(\mu'^2,\msusy^2)\,. 
\end{equation}

By using Eq.~\refeq{extrema} one can derive
\begin{equation}
\left(m_{S}^2 +m_S'^2 +\mu'^2\right)\gtrsim\max(\mu'^2,\msusy^2)\,,
\end{equation}
which is trivially satisfied when $m_{S}'^2>0$ and $m_{S}^2<0$, as is the case at hand (see Appendix~\ref{softs}).
On the other hand, requiring tree-level stability of the pseudo-scalar masses, whose expressions can also be found in~\cite{Ellwanger:2009dp}, 
yields an additional constraint: 
\begin{equation}
-2m_{S}'^2-\frac{\left(\xi_S+\xi_F\mu'\right)}{s}>0\,.
\end{equation}
Thus Eq.~\refeq{extrema} also implies $(m_{S}^2 -m_{S}'^2 +\mu'{}^2)>0$, which in turns implies $m_{S}'^2<\mu'{}^2$\,.
Assuming a relation $m_{S}'^2=-c \mu'\lsusy$, with $c>0$, just like in Eq.~\refeq{eff_model}, one gets
\beq\label{ineq}
-\mu'\gtrsim c\,\La\,.
\eeq

These arguments show that for small \tanb, which is the case of interest here, one can get reasonable physics only when the mass scale 
associated with $\xi_S$, $\xi_F$, $m_S'^2$, and $\mu'$ is much larger than \msusy. 

By reinserting Eq.~\refeq{ineq} back into Eqs.~\refeq{ext}-\refeq{extrema} one can finally infer that 
$ s + \xi_F/\mu'\approx 0$ and $s+\xi_S/m_{S}'^2\approx 0$ should simultaneously hold, up to terms that are much smaller than \msusy. 
Thus, one can derive the relation \refeq{ratio},
\beq\label{ratio2}
\frac{\xi_S}{\xi_F}=\frac{m_S'^2}{\mu'}\,,
\eeq
which must hold at the EW scale.

To summarize, we present all the coefficients for which the hierarchy $\msusy\ll \lsusy\lesssim |\mu'|$ holds 
in terms of $\msusy$, $\lsusy$, and $|\mu'|$:
\beq\label{estim}
\xi_F\approx -\msusy \mu',\quad \xi_S\approx \msusy\lsusy \mu',\quad m_S'^2\approx -\lsusy \mu'
\eeq
Our numerical analysis confirms these predictions. 
Note that $\mu'<0$ implies $\xi_F>0$. The sign of $\xi_F$ (as well as that of $\la$) can be always be chosen positive 
by rescaling the $S$ and, e.g., $H_u$ superfields.

As we claimed in \refsec{sec:eff}, the relation \refeq{ratio2} is set at the \my scale when appropriate instanton corrections are chosen.  
More importantly for the fine tuning calculation, however, Eq.~(\ref{ratio2}) is approximately RG invariant, so that it continues to hold at different scales.
We prove this by writing down the one-loop leading terms in the RG running:
\beq\barr{ll}
\d \log\xi_F=\la^2,&\quad\d \log\mu'=2\la^2\\
\d \log\xi_S\approx\la^2(1+2A_\la\xi_F/\xi_S),&\quad \d \log m_S'^2=2\la^2(1+2A_\la\mu'/m_S'^2)\,,
\earr
\nn
\eeq
where we have used the estimates in \refeq{estim} and the simplified notation $\d=16\pi^2 \frac{d}{dt}$.
The terms in parentheses are approximately 1 so that
\beq
\d\log\left( \frac{\xi_S}{\xi_F}\right)\approx\d\log\left(\frac{m_S'^2}{\mu'}\right)\approx 0\,,
\eeq
which leads to Eq.~\refeq{ratio2}.


\section{Tree level Higgs mass in the PQ NMSSM}
\label{app:higgs}

Tree-level masses of the three CP-even Higgs bosons in the PQ NMSSM can be calculated by diagonalizing the mass matrix explicitly given, e.g., 
in Ref.~\cite{Ellwanger:2009dp}. 
We give here approximate formulas for the lightest eigenvalue.
 
We assume the following hierarchy of the mass matrix eigenvalues: $m_{h_1}\ll m_{h_2}\ll m_{h_3}$\,. 
One can thus show that 
\beqa\label{higgs:tree}
m^2_{h_1,\textrm{tree}}\approx v^2\left(g^2\cos^2 2\beta+\lam^2\sin^2 2\beta\right)
+\frac{s v^2\lam^2(2s\lam-A_\lam\sin 2\beta-\mu'\sin 2\beta)^2}{\xi_S+\xi_F\mu'}\,.\label{step1}
\eeqa

Since $\mu'$ is much larger than the GMSB-generated soft terms given in Eqs.~\refeq{gmsbsoft}, one can further reduce Eq.~\refeq{step1} to
\beq\label{mhtree}
m_{h_1,\textrm{tree}}^2\approx M_Z^2\cos^2 2\beta+\lambda^2 v^2 \sin^2 2\beta \left(1+\frac{\mu'^2}{(\xi_S+\xi_F\mu')/s}\right)\,.
\eeq
Note that in the model presented in this paper $(\xi_S+\xi_F\mu')/s<0$\,. 

One can schematically separate Eq.~\refeq{mhtree} into the MSSM and \lam-dependent parts: 
$m_{h_1,\textrm{tree}}^2=M_Z^2 \cos^2 2\beta+\delta m_{h_1,\textrm{tree}}^2$\,.
One should recall from Appendix~\ref{estimates} that $\xi_F\approx -\mu's$\,, so that we get
\begin{equation}\label{happr}
\delta m_{h_1,\textrm{tree}}^2=\lambda^2 v^2 \sin^2 2\beta \left(1-\frac{1}{1+\frac{\xi_S}{\xi_F\mu'}}\right)\approx \lambda^2 v^2 \sin^2 2\beta \cdot \frac{\xi_S}{\xi_F\mu'}\,,
\end{equation}
which was used in the discussion of \refsec{sec:pheno}. 
As was mentioned over there, when $\xi_S\ll \xi_F\mu'$ one gets $\delta m_{h_1,\textrm{tree}}^2\approx 0$\,,
even for large values of $\lam\gtrsim 0.7$, which is the case presented in BP2 of Table~\ref{tab:benchmarks}.

Recall, finally, from Appendices~\ref{app:eff} and \ref{estimates} that $-\xi_S=2\Lambda\xi_F$\,, and 
$2\Lambda\lesssim -\mu'$\,, which implies that in our model $\epsilon=\xi_S/\xi_F\mu'\lesssim 1$.
As we discussed in \refsec{sec:pheno}, we found no points for which $m_{h_1,\textrm{tree}}^2> M_Z^2$\,,
so that large radiative corrections to the Higgs mass are always necessary 
to obtain $m_{h_1}\approx 125\gev$.

\bigskip

\bibliographystyle{JHEP}

\bibliography{KPS}

\end{document}